# Developing generative AI chatbots conceptual framework for higher education

Joshua Ebere Chukwuere

Department of Information Systems, North-West University, South Africa

joshchukwuere@gmail.com

**Abstract**

This research explores the quickly changing field of generative artificial intelligence (GAI) chatbots in higher education, an industry that is undergoing major technological changes. AI chatbots, such as ChatGPT, HuggingChat, and Google Bard, are becoming more and more common in a variety of sectors, including education. Their acceptance is still in its early phases, with a variety of prospects and obstacles. However, their potential in higher education is particularly noteworthy, providing lecturers and students with affordable, individualized support. Creating a comprehensive framework to aid the usage of generative AI chatbots in higher education institutions (HEIs) is the aim of this project. The Generative AI Chatbots Acceptance Model (GAICAM) is the result of this study's synthesis of elements from well-known frameworks, including the TAM, UTAUT2, TPB, and others along with variables like optimism, innovativeness, discomfort, insecurity, and others. Using a research method that encompasses a comprehensive analysis of extant literature from databases such as IEEE, ACM, ScienceDirect, and Google Scholar, the study aims to comprehend the implications of AI Chatbots on higher education and pinpoint critical elements for their efficacious implementation. Peer-reviewed English-language publications published between 2020 and 2023 with a focus on the use of AI chatbots in higher education were the main focus of the search criteria. The results demonstrate how much AI chatbots can do to improve student engagement, streamline the educational process, and support administrative and research duties. But there are also clear difficulties, such as unfavorable student sentiments, doubts about the veracity of material produced by AI, and unease and nervousness with new technologies.

## Introduction

Artificial intelligence (AI) chatbots in recent years have been used across disciplines (Yang & Evans, 2019). Generative AI Chatbots such as ChatGPT, HuggingChat, Google Bard, Microsoft Bing AI, Zapier AI Chatbot, and many others are becoming a norm across industries like journalism, content creation, health, finance, the retail sector, and the education sector. The emergence of AI chatbots in higher education (HE) provides opportunities as well as challenges for educators, governing bodies, and policymakers. The potential of AI Chatbots in HE is seen in their ability to assist many students cost-effectively (Momonov & Mirtskhulava, 2021; Sandu & Gide, 2019). However, the generative AI chatbots in education are still a shock to many because of their potential to revolutionize the system and provide quick answers to educational questions, explanations, and further viewpoints (Labadze, Grigolia & Machaidze, 2023).

Generative AI chatbots (AI chatbots) can improve learner engagement in HE by providing personalized assistance to students in advanced and specialized subjects (Ilieva, Yankova, Klisarova-Belcheva, Dimitrov, Bratkov & Angelov, 2023). Generative AI chatbots can also help with research-related tasks for lecturers and students, such as literature reviews and advice on research methodology, assist with homework and administrative tasks, reduce the workload of lecturers, and provide real-time feedback to lecturers and students. Furthermore, research carried out at Plovdiv University Paisii Hilendarski shows that a significant portion of students are aware of and have already utilised the instructional potential of these newly developed AI chatbots (Ilieva et al., 2023). The researchers further found that a sizable majority of students indicated a strong desire to utilise AI chatbots and reported great happiness with these technologies. According to Momonov and Mirtskhulava (2021), the potential adoption of AI chatbots in HE is increasing but more research is needed. The revolution of AI chatbots in HE is still at an early stage and many are yet to adopt it, then more research ideas are necessary to assist in the implementation process. Then, this study aims to propose a framework for the adoption of AI Chatbots in higher education.

**Background of the AI chatbots in HE**

Chatbots can be regarded as computer algorithms with the ability to understand human intents, and imitate human information and conversation (Iiieva et al., 2023). The ability of chatbots involves the provision of personalized real-time responses to questions and queries. This computer algorithm functions as a result of its ability to learn and re-learn from a large dataset and form new ideas. Recently, the number of AI chatbots kept increasing, and investment by multinational companies like Amazon Lex, IBM Watson Assistant, Openai, ERNIE Bot, Google Bard, ChatGLM-6B, PanGu-Bot, Yandex (YaLM) Chatbot, Alpaca, BLOOMChat, Microsoft Bing and Edge Chat, ChatGPT Plus, and Vicuna to provides conservational using machine learning and AI algorithm to mimic human interactions (Iiieva et al., 2023). These AI chatbots are currently used for educational purposes like teaching and learning pedagogy, research, innovations, and non-educational purposes.

According to Ilieva et al. (2023), the late 1960s saw attention from educational institutions investing in chatbots to mimic human interaction for language teaching, then, in the 21$^{st}$ century, there was an increase in conversational AI and embedment of AI chatbots in conventional electronic learning systems. The AI chatbots as a pedagogical tool have the following features: conversational assistance, multi-modality, multilingual support, cost-effectiveness, integration with other software systems, and data analytics and insights to support lecturers and students teaching and learning experience (Ilieva et al., 2023).

**Literature review**

**AI Chatbots in higher education an overview**

The advancement of technology in today's world impacts every aspect of human society. The use of virtual assistants (chatbots) is changing the conventional ways of interaction with the computer system. According to Gupta, Hathwar and Vijayakumar (2020), a chatbot is a software program that uses natural language understanding and processing to communicate with humans through interactive queries. It acts as a virtual assistant that can complete tasks, provide entertainment, and offer business strategy tips. Generative AI Chatbots simulate the cognitive abilities of humans and use technologies such as artificial intelligence, machine learning, neural networks, and natural language processing to understand and respond to user queries. Gupta et al. (2020) divide chatbots into three categories according to the simplicity of their underlying technology, algorithms, and user interfaces:

1. **Menu/button-based chatbots:** These are the most popular and basic kinds of chatbots. To help the user navigate a decision tree and get the right AI response, they employ buttons and top-down menus. These chatbots can't always be relied upon to provide the correct response, though, as their performance is often slower.

2. **Keyword recognition-based chatbots:** These chatbots use natural language processing to recognize keywords in the user's query and provide a relevant response. They are more flexible than menu-based chatbots and can handle a wider range of queries. However, they may not be able to understand complex queries or respond accurately to misspelled or ambiguous words.

3. **AI/NLP-based chatbots:** These chatbots use advanced technologies such as machine learning, neural networks, artificial intelligence, and natural language processing to simulate human-like conversations with users. They can understand complex queries, learn from user interactions, and provide personalized responses. However, they require a large amount of data and training to achieve high accuracy and may be more expensive to develop and maintain. The above classifications are increasingly used in higher education especially the AI/NLP-based chatbots such as ChatGPT, Google Bard, and many others which are revolutionizing the sector.

The use of generative AI chatbots is advancing the interest of lecturers and students in adding value to higher education teaching and learning process in the following manner: personalized assistance (meeting specific learning needs), independent and self-directed learning, and scholarly resource access (assist in access to the literature review and research methodologies) (Ilieva et al., 2023).

**AI chatbots concerns**

The application of AI chatbots in higher education is drawing concerns (Chan, 2023). The biggest concern among scholars is that students will use AI chatbots to cheat, and plagiarise their assignments, research, and other academic task (Chan, 2023). Academic cheating has been long in the history of academics. The act happens at every level of education, from primary to tertiary involving misconduct (Barbaranelli, Farnese, Tramontano, Fida, Ghezzi, Paciello & Long, 2018). Academic cheating is a violation of academic integrity. Academic integrity is built on five fundamental pillars and values "honesty, trust, fairness, respect, and responsibility" (Part, Part, Jang & 2013). Academic cheating is regarded as misconduct committed by a student to deceive his or her lecturer into believing that a submitted academic assignment is own work (Dejene, 2021). The act involves deceiving, tricking, misleading, fooling, or defrauding another intentionally.

Scholars are concerned about ethical issues on the biases of AI-generated text and content. According to Kooli (2023), AI chatbots are as biased as the data it was trained on, if the data entered is biased, then quire outcome will be biased, discriminatory, and inequality. Security is another aspect of concern for many because of the conservations that happened on AI chatbots and the storage. However, Hasal, Nowaková, Ahmed Saghair, Abdulla, Snášel and Ogiela (2021) believe that security issue is not a major concern for AI chatbots because many of their security concerns are known and mitigation mechanisms are available.

**Theoretical frameworks**

Many theoretical frameworks are applied in carrying out information systems (IS) research. The section seeks to understand some existing frameworks that can contribute to the implementation of generative AI chatbots in HE. According to Ilieva et al. (2013), scholars are beginning to explore frameworks and ways to assist the execution of AI chatbots in HE. For example, Rasul et al. (2023) through constructivism learning theory presented a framework to inform the usage of ChatGPT in universities. Gimpel et al. (2023) provided a step-by-step on how lecturers and students can use ChatGPT to improve teaching and learning practice (Su & Yang, 2023). Chan (2023) proposed a framework to address ethical, privacy, security, and accountability issues. While Ilieva et al. (2023) proposed a framework to advance the resources offered in teaching and learning in higher education. These proposed frameworks provide different aspects of the application of AI chatbots to positively impact on current teaching and learning process in higher education institutions. However, according to Ilieva et al. (2023), there are drawbacks in frameworks because they failed to address:

**a.** Syllabus operational issues on knowledge acquisition and skills development.

**b.** The usability analysis of AI chatbots focused on AI tool's implications in higher education.

**c.** The readiness and perception of the students, lecturers, and other stakeholders were neglected but focused on the classroom process and content provided.

**Problem statement**

AI chatbots have the potential to transform pedagogical activities and reshape the educational landscape (Ilieva et al., 2023). However, the adoption rate of AI chatbots in HE is still low (Yang & Evans, 2019). The adoption of AI chatbots in HE is not without its challenges. One challenge is that some students may express a negative attitude towards chatbots, stressing the value of in-person collaboration in the development of their soft skills (Ilieva et al., 2023). Finally, some students may have concerns about the authenticity of the texts generated by AI chatbots, which can be addressed by verifying them through AI technology like ChatGPT (Ilieva et al., 2023).

Scholars continue to explore the adoption of AI chatbots in HE (Ilieva et al., 2023), the ability to improve students learning (Momonov & Mirtskhulava, 2021), and the benefits and challenges (Okonkwo & Ade-Ibijola, 2021). AI chatbots in higher education research and publications have been growing in the past few years (Momonov & Mirtskhulava, 2021). The growing attention on AI chatbots is making scholars research and propose mechanisms to strengthen the usage and application of AI chatbots in higher education. Ilieva et al. (2023) proposed a theoretical framework to enable students to attract effectively with generative AI tools in a blended learning process. The framework failed to guide the adoption of this emerging technology in HE settings. However, scholars are advancing in their quest to provide a mechanism for making AI chatbot usage in higher education settings. Despite the growing, and availability of academic research and knowledge, there is a lack of a comprehensive approach that informs the initialization, analysis, design, and implementation of robust adoption of AI chatbots in HE (Ilieva et al., 2013). However, the presence of a comprehensive framework for HEIs in the adoption of generative AI chatbots is missing and limited in academic published literature at the time of the study. Then, this study combines various components from widely and less-known frameworks used in information systems research and other social science research to propose a comprehensive framework to guide the adoption of generative AI chatbots in HE.

**Research objectives and questions**

The purpose of this study is to provide a framework for the adoption of generative AI chatbots in HE. The framework outlines the variables (factors) that affect HEIs adoption of AI chatbots. According to this study, GAICAM is a strong framework that can help universities successfully use generative AI chatbots. In order to handle the possibilities as well as the constraints of generative AI chatbots in HE, this framework strikes a balance between technological preparedness and user attitudes. This study adds to the expanding corpus of research on artificial intelligence applications in education and offers educational institutions a strategic roadmap for incorporating generative AI chatbots into their systems.

**Research methodology**

There are many research methods that can be used to explore the acceptance of generative AI chatbots in higher education. Ilieva et al. (2023) and Sandu and Gide (2019) used a survey involving university students. Ou, Stöhr and Malmström (n.d) used post-humanist in qualitative research among university students. While Momonov and Mirtskhulava (2021) used secondary data in reviewing empirical literature studies. This research was conducted by searching online databases used for information systems, social science, and education research: Association for Computing Machinery (ACM) digital library, Institute of Electrical and Electronics Engineers (IEEE), ScienceDirect, Google Scholar, and ResearchGate.

*Terms search*

The acceptance of AI chatbots in HE is emerging and literature is growing (Momonov & Mirtskhulava, 2021). Chatbots were the main searched item for understanding its application by students, lecturers, and higher education at large. The study searched terms like "chatbots for higher education", "AI chatbots effect in higher education", "AI chatbots framework for higher education", "AI chatbots use by students", "AI chatbots usage by lecturers", and "AI chatbots".

*Criteria for inclusion in the study*

The articles included in the study are based on 1) written in the English language, 2) peer-reviewed academic materials, 3) empirical studies focusing on the use of AI chatbots in HE, and articles published between 2020 and 2023. This review study is informed by the following research questions:

1. How can AI chatbots affect HE?
2. What are the frameworks to aid the adoption of generative AI chatbots in HE?
3. What are the components to consider in developing a generative AI chatbot framework for HE?

**Discussion of the findings**

**AI chatbots usage in HE**

The use of AI chatbots in HE is growing. A study found that AI chatbots can be an effective tool for providing quick and accurate responses to student queries, reducing the workload of university staff, and improving student satisfaction (Gupta et al., 2020). Sandu and Gide (2019) also found that AI chatbots enhance communication, productivity, and efficient teaching and learning assistance. The AI chatbots further make higher education students' learning experience customizable, answer their questions and curiosities, and advance their learning abilities and skills (Liu, Subbareddy & Raghavendra, 2022). The thinking ability needs and expectations of students and academicians in higher education are improved through AI chatbots in making learning fun and interesting in providing access to innovative knowledge and skills.

**The frameworks to aid the adoption of generative AI chatbots in HE**

The use of information systems in addressing social issues and needs keeps growing. The application involves the use of theoretical and conceptual frameworks in the process. According to Chukwuere (2021) and Adom, Hussein and Agyem (2018), a theoretical framework is the base upon which academic ideas are built in addressing research objectives and questions. A conceptual framework is an idea built on theoretical framework components. Several theoretical frameworks are often applied in IS research. This section of the study briefly discussed the existing theoretical frameworks and drew necessary components that were applied in formulating the proposed AI chatbots conceptual framework called Generative AI Chatbots Acceptance Model (GAICAM) as discussed below.

a. **Theory of Reasonable Action (TRA):** TRA, which was created by Ajzen and Fishbein in 1975, is a popular theory for forecasting and understanding behaviour in people. According to this theory, people's attitudes and subjective standards around particular human behaviours in an issue influence their behavioural intention (Alkhwaldi & Kamala, 2017). TRA has been criticised, meanwhile, for being a broad model and failing to take into account additional factors that affect people's intentions and behaviours (Alkhwaldi & Kamala, 2017). The model is among the most used model to understand one's motivational factors, behavoural intention, and attitude regarding actual behavoural action taken by someone (Chukwuere, Ntseme & Shaikh, 2021). In the context of technological adoption and usage, one's motivational factors, behavioural intention triggers actual behavioual action towards adoption. The theory looks into the association between "belief, intention, attitude, and behaviour", in believing that human actions are determined by available implications (Tlou, 2009). The theory was able to understand the behavioral intention in a given action but failed to acknowledge some underlying issues such as innovations, cost, and many more.

b. **Theory of Planned Behavour (TPB):** TPB, which Ajzen developed in 1991, is a development of TRA that incorporates the perceived behavioural control construct. According to TPB, an individual's attitudes, subjective norms, and perceived behavioural control, all influence their behavioural intention. TPB is frequently used to forecast and interpret behaviour in people in a variety of settings (Alkhwaldi & Kamala, 2017). The theory outlines the conditions under which an individual's behavioural purpose towards attitude is set off. One of the most well-known theoretical frameworks for comprehending and forecasting human behaviour is the Theory of Planned Behaviour (TPB). It suggests that three primary factors—an individual's attitude towards the behaviour, subjective standards, and perceived behavioural control—have an impact on their intention to carry out a behaviour. Subjective norms capture the sense of felt social pressure to engage in the behaviour, attitude expresses how the person views the behaviour, and perceived behavioural control includes the sense of how easy or difficult the behaviour is to accomplish. TPB has been used in a variety of contexts and has shown to be useful in figuring out people's intents to utilise and accept technology, among other things.

c. **Technology readiness index (TRI 2.0):** This is an offshoot of the technology readiness index (TRI 1.0) with a focus on providing enabling factors that position an individual's mindset in acceptance of new technology to achieve the personal or organizational task (Ntseme, 2019; Parasuraman, 2000). On TRI 2.0, optimism and innovativeness are technology readiness contributors/enablers while discomfort and insecurity are the technology readiness inhibitors (Ntseme, 2019; Parasuraman, 2000). For example, individuals with high technology readiness index regard the benefits of technology which define their behaviour towards it.

d. **Technology Acceptance Model (TAM):** The Technology Adoption Model (TAM), which was developed by Davis et al. in 1989, asserts that two particular beliefs—"perceived usefulness" and "perceived ease of use"—are the most major drivers for the actual behaviour of technology adoption from the perspective of the individual. The Technology Acceptance Model (TAM) is a theoretical framework that examines and analyses aspects that influence the acceptability of information technology in order to explain and forecast its acceptance (Liao, Hong, Wen & Pan, 2018). Numerous scholars have applied and expanded the TAM model, which is a dependable, strong, and well-established theory (Alkhwaldi & Kamala, 2017). It provides the basis to determine if information systems artifacts will be accepted and used by individuals voluntarily (Ntseme, 2019). The framework uses "perceived usefulness (PU)", "perceived ease of use (PEU)", "attitude towards using (ATU)", and behavioural intention as factors to determine individual or organizational adoption of information systems in practice.

According to the study, a theoretical framework that describes how people come to embrace and use new technology is called the TAM. Since its first proposal by Fred Davis in 1989, the model has found widespread use in the field of information systems research. According to the TAM, perceived utility and perceived ease of use are the two primary aspects that impact consumers' attitudes toward technology. The degree to which a user feels that technology will enable them to carry out their activities more successfully is known as perceived usefulness, whereas the degree to which a user believes that technology is simple to use is known as perceived ease of use. These two elements have a direct impact on users' intentions to utilise a technology, which in turn affects how the technology is used, according to the TAM. According to the paradigm, users' attitudes towards technology can also be indirectly influenced by external variables that affect their views of its utility and ease of use, such as social influence and enabling conditions. The TAM has been extensively utilised in information systems research to assess the acceptance and utilisation of diverse technologies. It offers a valuable framework for comprehending how users learn to accept and employ new technologies.

e. The **Unified Theory of Acceptance and Use of Technology (UTAUT):** A comprehensive model that aims to describe and forecast people's adoption and use of technology is called the Unified Theory of Adoption and Use of Technology (UTAUT). UTAUT, which was created in 2003 by Venkatesh et al., incorporates components from eight well-known models of technology adoption, such as the Motivational Model (MM), the Theory of Reasoned Action (TRA), and the Technology Adoption Model (TAM). Performance expectancy, which measures how much a person believes using a technology will improve their job performance, effort expectancy, which measures how easily a technology can be used, social influence, which measures the effect of social factors on a person's decision to use technology, and facilitating conditions, which measures how much a person believes the organisational and technical infrastructure supports the technology use, are the four main constructs identified by UTAUT as influencing technology acceptance and use. Furthermore, UTAUT takes into account moderating variables including age, gender, experience, and voluntariness of usage, all of which might affect how the major conceptions and acceptance of technology relate to one another. All things considered, UTAUT offers a thorough framework for comprehending the elements that affect people's acceptance and use of technology, and it has been extensively employed in studies to evaluate technology adoption and usage in a variety of circumstances.

f. The **Motivational Model (MM)** aims to explicate individuals' motivation to use technology. Developed by Venkatesh in 2000, MM is a model that explains the factors that influence individuals' motivation to use new technologies. MM includes three constructs: intrinsic motivation, extrinsic motivation, and self-efficacy. MM has been used in various studies to predict and explain individuals' motivation to use new technologies (Alkhwaldi & Kamala, 2017). The framework, which Davis developed in 1992, contends that self-efficacy, extrinsic incentive, and intrinsic motivation are the three primary determinants of people's drive to utilise technology. Extrinsic motivation is influenced by outside forces like incentives or peer pressure, whereas intrinsic motivation is the person's inherent desire to utilise technology. Self-efficacy is a measure of a person's confidence in their capacity to use technology efficiently. The MM has been applied in the study to understand the underlying motivations driving individuals' adoption and use of technology, providing insights into the factors that influence users' motivation and subsequent technology usage.

g. **Unified Theory of Acceptance and Use of Technology 2 (UTAUT2):** Venkatesh et al. created UTAUT2, an expanded version of UTAUT, in 2012. With a strong predictive power of 74% of the variation in usage behavioural intention, UTAUT2 comprises nine IT acceptance models. Understanding the measurements of human acceptance or rejection behaviours towards new technologies is made possible by UTAUT2, a dynamic and comprehensive theoretical model that may take into account cultural, social, technological, and other relevant behavioural determinants (Alkhwaldi & Kamala, 2017).

h. **Revised Technology Acceptance Model (RTAM):** The study's revised TAM aims to encompass and depict the cultural elements that impact human preferences in many social situations, including developing and rising nations. Recognising that technology is utilised and deployed in an environment governed by cultural, social, economic, and political elements, the model seeks to construct a reflective system based on individual choices and social and cultural values (Chukwuere et al., 2021).

The five layers that make up the revised TAM are designed to separate and categorise the model into separate but related structures. The first layer consists of the enablers that shape middle-class and developing nations' user attitudes, customs, and cultures around technology. Perceptions mentioned by TAM and other elements absent from current theories, models, and frameworks that specify the specific characteristics influencing technology adoption in developing and emerging nations comprise the second layer. The layers that follow deal with things like infrastructure, belief systems, real usage, behaviour, attitude, perception, and culture/tradition.

The model highlights how Social, Economic, and Political (SEP) aspects play a major role in determining how technology is adopted in developing and rising nations. It acknowledges that people's views and behaviours about the adoption of technology are greatly influenced by sociological, cultural, political, and economic factors. With a focus on developing and emerging nations in particular, the updated TAM seeks to offer a more thorough framework for comprehending and encouraging technology adoption in a variety of social circumstances.

**The components to consider in developing an AI Chatbots framework for higher education**

The Generative AI Chatbots Acceptance Model (GAICAM), Figure 1 is composed of different framework components such as TAM, TRI 2.0, UTAUT2, revised TAM, and TPB.

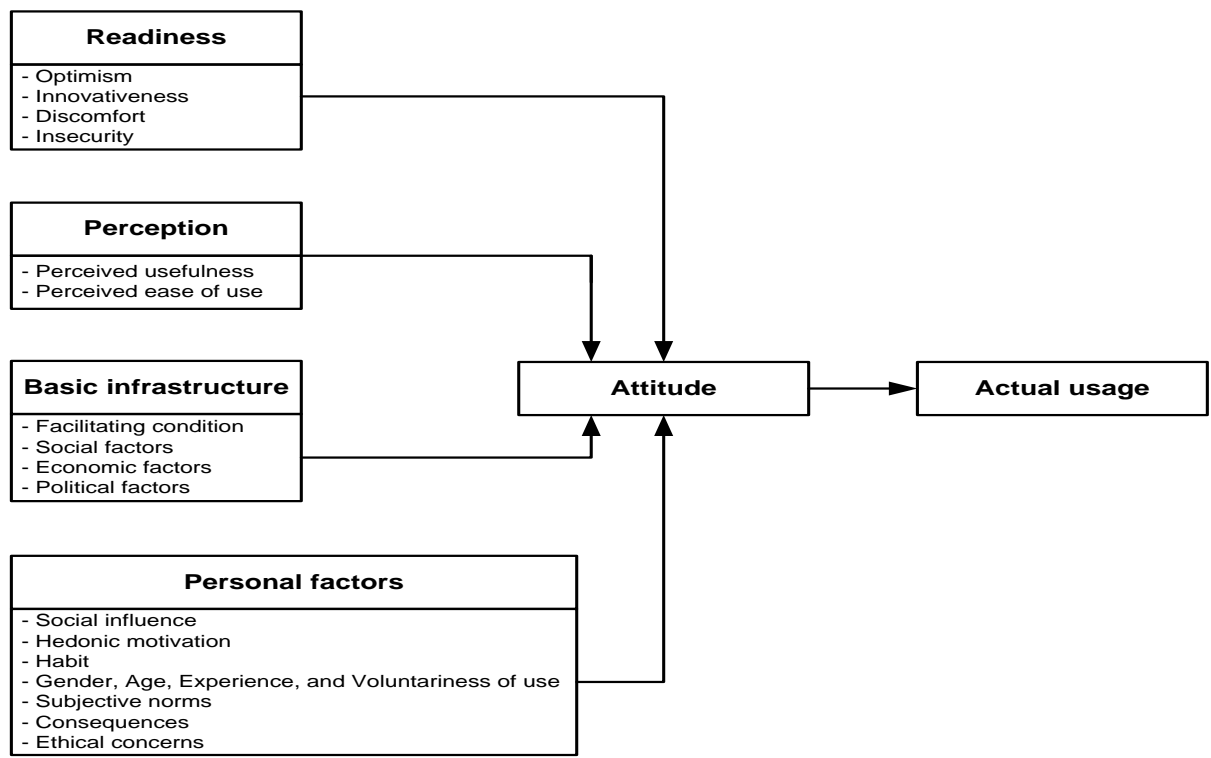

**Figure 1:** Generative AI Chatbots Acceptance Model (GAICAM)

**<u>Readiness</u>**

Readiness provides the aspect of the user's willingness to adopt technology and generative AI chatbots in particular. The readiness outlined and discussed the components that enable the propensity to use generative AI chatbots in higher education.

**Optimism:** It provides individuals "positive view" towards increasing life and work efficiency (Ntseme, 2019; Parasuraman, 2000). Through optimism, technology users have total control of the new technology with the belief that it adds value to their tasks. For example, optimist users will likely use the new technology more often to achieve basic tasks and functions. This study, optimism shows a positive view about AI chatbots with the belief that it will offer increased control, flexibility, and efficiency in daily academic and non-academic tasks. In the context of AI chatbots in HE, optimism would manifest as a positive belief in the potential of AI chatbots to improve the learning outcome and experience. This includes the perception that generative AI chatbots can provide personalized support, improve accessibility to educational resources, and contribute to a more efficient and effective learning process. The study believes that optimism reflects a positive outlook on the potential benefits of generative AI chatbot technology in facilitating the process of teaching and learning, and research.

**Innovativeness:** according to Ntseme (2019) and Parasuraman (2000), innovativeness provides the extent one believes that he/she is a technology thought leader in trying out new technology. An individual trying new technology in the case of an AI chatbot means it's innovative and interesting, the two features provide the basis for one to try it. Innovativeness, as described by Parasuraman (2000), reflects an individual's inclination to be a pioneer and thought leader in embracing new technology. This trait encompasses a willingness to explore and adopt technology-based products and services, positioning the individual as a leader in technological advancements. Tsikriktsis (2004) further emphasizes that innovativeness looks at the extent to which people believe they are at the forefront of trying out new technology, demonstrating a proactive and forward-thinking approach. Son and Han (2011) also note that highly innovative individuals feel comfortable with technology and require less proof of its outcomes, showcasing a readiness to embrace and lead in the adoption of new technologies.

For example, for an generative AI chatbot to appeal to students, lecturers, and other stakeholders, it must be innovative which drives confidence towards acceptance and usage. To this study, innovativeness is regarded as one's ability to accept an generative AI chatbot based on its innovativeness to impact positively their academic and non-academic lives. Innovativeness in the context of generative AI chatbots in HE would involve a willingness to embrace and explore the potential of generative AI chatbots as educational tools. This includes being open to experimenting with new AI-driven learning methods, actively seeking out opportunities to integrate generative AI chatbots into educational practices, and being among the first to adopt and advocate for their use in educational settings. Innovativeness according to this study pertains to a willingness of the educational stakeholders to explore and adopt new AI technological advancements. For the application of AI chatbots in education, innovativeness would manifest as a tendency among students and lecturers to actively explore and champion the integration of generative AI chatbots as valuable educational tools, driving innovation and progress in educational practices.

**Discomfort:** This means individuals believe that they have no control over new technology (Ntseme, 2019; Parasuraman, 2000). They believe that new technology is overwhelming, and distrusting. In the case of the application of generative AI chatbot in higher education, discomfort individuals will consider it not important because of their belief that the technology is overwhelming and difficult to understand. The discomfort is a result of negative beliefs and perceptions of the technology. According to this study, discomfort means a negative feeling towards generative AI chatbots because they believe that it is overwhelming and they cannot control the application in their academics. For the generative AI chatbot to be implemented in higher education, then the issue surrounding discomfort must be addressed to guarantee acceptance. Discomfort may arise from concerns about the integration of generative AI chatbots in HE, such as fears of job displacement for educators, ethical considerations related to data privacy and security, or uncertainties about the impact of generative AI chatbots on traditional teaching methods. Overcoming discomfort involves addressing these concerns through transparent communication, ethical guidelines, and professional development opportunities for educators. Discomfort encompasses concerns and unease about the integration of new technologies in the case of generative AI chatbot adoption in HE.

**Insecurity:** Insecurity, as it pertains to technology adoption, encompasses individuals' doubts and skepticism regarding the reliability and security of new technologies. This includes concerns about the potential for technical malfunctions, data privacy, and security issues, and the overall trustworthiness of the technology in fulfilling its intended functions. According to Parasuraman (2000), insecurity is a significant inhibitor to technology readiness, as it can lead individuals to question the dependability of technology in meeting their needs. Additionally, Tsikriktsis (2004) emphasizes that insecurity is linked to discomfort but focuses specifically on doubts about the functionality and reliability of technology. Addressing these insecurities is crucial for fostering trust and confidence in the adoption of new technologies, ensuring that individuals feel secure in their interactions with technology and are willing to embrace its potential benefits (Ntseme, 2019).

According to this study, insecurity relates to doubts about the reliability and security of new technologies. In the context of generative AI chatbots in HE, insecurity may stem from doubts about the reliability and effectiveness of generative AI chatbots in supporting learning learning and outcomes. This includes concerns about the accuracy of AI-generated educational content, the potential for biases in AI algorithms, and the overall trustworthiness of AI chatbot interactions. Addressing insecurity involves ensuring transparency in generative AI chatbot functionalities, providing evidence of their educational benefits, and implementing mechanisms to mitigate biases and errors.

## **Perception**

User's acceptance of any technology is defined by their perception of the potential benefits to derive. The perception of the GAICAM is defined by the factors discussed here.

**Perceived usefulness:** The degree to which an individual feels that utilising a certain technology will improve their ability to accomplish their work is known as perceived usefulness. It represents the user's subjective likelihood that utilising the system would be advantageous, making it a crucial factor in determining an individual's desire to utilise technology (Alkhwaldi & Kamala, 2017). A user's subjective likelihood that utilising a technology or system would improve their work performance is known as their perceived utility (PU) (Amadu, Muhammad, Mohammed, Owusu & Lukman, 2018). PU captures students', lecturers, institution management, and other stakeholders' beliefs about the extent to which AI chatbots can enhance their academic performance. If students and other stakeholders perceive generative AI chatbots as valuable tools for effective information search, exploration of ideas, learning, quick access to learning and research content, exchanging information, interacting with peers, and improving their learning outcomes, they are more likely to utilize it for learning purposes.

**Perceived ease of use (PEOU):** The degree to which a person thinks utilising a specific technology would be effortless is known as perceived ease of use. According to Amadu et al. (2018), PEOU is the extent to which a user anticipates a system or technology to be effortless to use. This part focuses on how easy it is for the user to utilise the system and how easy they think it is to use. Perceived utility and perceived ease of use are key concepts in the TAM because they influence users' attitudes and behavioural intentions toward embracing new technologies in important ways (Alkhwaldi & Kamala, 2017). Overall, PEOU and PU play crucial roles in shaping students', lecturers', institution management, and other stakeholders' attitudes and behaviors toward using generative AI chatbots for teaching and learning, and research, highlighting the importance of addressing these factors when implementing educational initiatives involving generative AI chatbots in higher education.

## Basic infrastructure

The basic infrastructure provides the basic resources that facilitate the use of generative AI chatbots in higher education. The infrastructure involves facilitating conditions, and economic and political factors.

**Facilitating conditions:** The user's impression of the tools and assistance available to carry out an action is referred to in this dimension. It reflects the user's perception of the system's compatibility with its existing infrastructure and the availability of technical support (Alkhwaldi & Kamala, 2017). These are objective factors like external factors to individuals as an enabler towards acceptance and non-acceptance of technology (Maruping, Bala, Venkatesh & Brown, 2017). Marikyan and Papagiannidis (2021) define this dimension as the extent to which a person has access to the tools and assistance required to use a certain technology efficiently. It influences the actual utilization of technology and is an important factor in the UTAUT model (Venkatesh, Thong & Xu, 2016). As a dimension that focuses on resource perception, then, the adoption of generative AI chatbots in HE is centered on the available resources such as technical and non-technical resources. Lecturers, institutions, and other stakeholders should make resources to aid the adoption of generative AI chatbots in higher education.

**Social factors:** Social factors encompass the societal and cultural influences that shape individuals' attitudes and behaviors towards technology adoption. The factor highlights the crucial determinants in technology acceptance processes, especially in developing and emerging countries. Chukwuere, Ntseme and Shaikh's (2021) study emphasizes that societal and natural changes within the environment directly affect the adoption of technology by individuals. It is noted that culture and its attributes play a significant role in guiding social factors in these countries. Therefore, understanding the social dynamics and cultural norms is essential for comprehending generative AI chatbots adoption patterns in these contexts.

**Economic factors:** Economic variables are defined as financial concerns that have a major influence on people's judgments about the adoption of technology. According to research, economic concerns go beyond financial factors that influence people's ability to afford new technology (Chukwuere et al., 2021). It is emphasized that financial crises can lower people's ability to spend, which can affect how people embrace new technologies. Financial limits and constraints are important factors that influence the adoption and use of generative AI chatbots in HE in underdeveloped and emerging nations.

**Political factors:** Political considerations are laws, regulations, and other acts of the government that affect how technology is adopted in a community (Chukwuere et al., 2021). It is significant to remember that government policies and governance are essential in fostering an atmosphere that supports the acquisition, use, and use of technology for social and economic advancement. It is well known that laws and regulations have an impact on the advancement and uptake of technology. This study underscores the importance of good governance in promoting a conducive environment for technology exploration and adoption of generative AI chatbots in higher education.

## Personal elements

Personal elements are those direct and indirect factors that define whether generative AI chatbots will be accepted in higher education.

**Social influence:** This dimension describes the degree to which users feel that significant others like family and friends think they ought to utilise a specific technology. It reflects the user's perception of the system's social norms and the influence of others on their decision to adopt the technology (Alkhwaldi & Kamala, 2017). While Maruping et al. (2017) believe that social influence focuses on the views of others on one using technology. The dimension describes how much a person is impacted by the beliefs and behaviours of other people, whether peers or superiors. According to this study, other students', lecturers', and other institutions' perspectives and opinions have an impact on the use of generative AI chatbots in HE. The more others use and adopt the technology for educational purposes, it influence others to join. This dimension is a fundamental component of the UTAUT model and has a significant influence on how people intend to utilise technology (Venkatesh et al., 2016).

**Hedonic motivation:** This component, which relates to the enjoyment or fun of utilising technology, has been demonstrated to be crucial in influencing the adoption and usage of that technology. It expresses how much the user enjoys using the system and how much of a good emotional experience it offers (Alkhwaldi & Kamala, 2017). According to Venkatesh et al. (2016), the UTAUT includes the concept of hedonic motivation, which describes how much pleasure or delight a person gets from utilising technology. It has been discovered that it plays a crucial role in influencing technology adoption and usage and is a major predictor of both (Brown & Venkatesh, 2005). The foundation of hedonic motivation is the idea that people are driven to utilise technology not just for its practical advantages but also for the happiness and pleasure it brings. This concept is especially important when it comes to educational technology since the perceived hedonic character of the result (like perceived enjoyment) is a strong predictor of using technology to support learning.

**Habit:** This element describes the degree to which learning causes people to do behaviours automatically. It displays the degree to which technology has gotten ingrained in the user's everyday life and how frequently they utilise it (Alkhwaldi & Kamala, 2017). Many in the educational institution are becoming technology savvy in aiding academic and non-academic purposes. Then, the habitual use of technology in higher education is influential in the adoption of generative AI chatbots in HE.

**Gender, Age, Experience, and Voluntariness of use:** These are moderating variables that influence the relationship between the components of UTAUT2 and technology acceptance behavior. They reflect the user's demographic characteristics and the extent to which they have a choice in using the technology (Alkhwaldi & Kamala, 2017). This factor moderates social influence (Maruping et al., 2017). This implies that the ability of anyone (students, lecturers, and other institutions) to adopt generative AI chatbots is influenced by the views and opinions of others surrendering him or her.

**Subjective norms:** Subjective norms are a UTAUT concept that describes the felt social pressure to participate in a specific behaviour, in this example, using technology (Marikyan & Papagiannidis, 2021). It represents a person's opinion on whether or not influential people believe they ought to utilise the technology. Since subjective norms reflect the impact of social circumstances on an individual's desire to use technology, they are a crucial concept in understanding technological acceptability and the usage of generative AI chatbots in higher education. This component refers to the social pressure that individuals perceive from significant others to perform or not perform a behavior (Alkhwaldi & Kamala, 2017). It reflects the individual's perceptual experience of the social norms surrounding the behavior and the extent to which they feel obligated to conform to these norms. Subjective norms are part of the social influence construct, which also includes other factors such as social factors and image. The influence of subjective norms on technology acceptance is moderated by gender, suggesting that the impact of social pressure on technology use may vary based on gender differences.

Understanding subjective norms is important for lecturers, education institutions, and other stakeholders, as it highlights the significance of social influence in shaping student's attitudes and intentions toward generative AI chatbots adoption. By recognizing the impact of subjective norms, higher education institutions can develop strategies to leverage social influence and promote positive attitudes towards generative AI chatbots use within higher education institutions.

**Consequences:** This component speaks to the results that people believe come from employing technology. It represents the person's opinions on the advantages and disadvantages of utilizing technology as well as how much they value these results. Although UTAUT does not specifically address repercussions as a fundamental concept, it does subtly recognize the possible results of adopting and using technology. The idea acknowledges that people's perceptions of the anticipated performance, effort involved, social pressures, and enabling environments all play a role in their decision to accept and utilize technology. These perceptions, in turn, can lead to various consequences, both positive and negative, as individuals engage with technology.

The consequences of generative AI chatbots adoption and use, such as increased productivity, improved access to academic resources, communication, social isolation, addiction, and privacy concerns, can be seen as outcomes that are influenced by the factors considered in UTAUT. For example, if a student, lecturers, and other stakeholders perceive high-performance expectancy and low effort expectancy from using particular generative AI chatbots in higher education, they may be more likely to adopt it, potentially leading to positive consequences such as increased productivity. Conversely, if social influence is a significant factor in generative AI chatbots acceptance in higher education, the consequences may include social pressure to adopt the technology, which could have both positive and negative implications. By understanding this factor, higher education institutions and researchers can gain insights into the potential outcomes of generative AI chatbots adoption and use, and develop strategies to promote positive consequences and mitigate negative ones.

**Ethical concerns:** The integration of generative generative AI chatbots in education presents a myriad of ethical concerns that necessitate careful consideration. One prominent issue revolves around data privacy and security, as AI chatbots can collect and store sensitive information, raising questions about data protection and confidentiality (Williams, 2024; Oniani, Hilsman, Peng, Poropatich, Pamplin, Legault & Wang, 2023). Ensuring compliance with data protection regulations and implementing robust measures to safeguard data from unauthorized access or misuse is imperative to uphold ethical standards and maintain trust in educational institutions. Furthermore, the potential for algorithmic bias in generative AI chatbots poses a significant ethical dilemma, as these systems may inadvertently perpetuate societal biases present in the data they are trained on, leading to unfair treatment or discrimination in educational interactions (Williams, 2024).

Another critical ethical concern is the risk of plagiarism facilitated by AI-generated content produced by AI chatbots, which may tempt students and scholars to present AI-generated work as their own, compromising academic integrity and devaluing the educational process (Yu, 2024; Williams, 2024; Miao, Thongprayoon, Suppadungsuk, Garcia Valencia, Qureshi & Cheungpasitporn, 2023). Addressing this issue requires a comprehensive approach, including the deployment of advanced plagiarism detection tools, clear policies on academic honesty, and innovative assessment methods that discourage unethical practices. By actively addressing these ethical challenges and promoting a culture of integrity and accountability, educators, policymakers, and students can harness the potential of chatbots in education while upholding ethical standards and fostering a supportive and inclusive learning environment (Williams, 2024; Akgun & Greenhow, 2022).

## Attitude

This component relates to a person's assessment of behaviour, whether it be favourable or bad. It represents the person's perceptions of the behavior's results and how much they are valued (Alkhwaldi & Kamala, 2017). According to the research by Amadu, Muhammad, Mohammed, Owusu and Lukman (2018), students' behavioural intention to use social media for collaborative learning is significantly shaped by the "attitude" component of the TAM. Perceived utility (PU) and perceived ease of use (PEOU) are two fundamental assumptions that impact attitude, according to TAM (Amadu et al., 2018). According to this study, students' attitudes towards utilising generative AI chatbots for academic purposes like completing research and handling class assignments are positively influenced if they believe these tools are user-friendly and can improve their academic performance.

As a mediating factor, attitude significantly influences students', and lecturers' intention to use generative AI chatbots for academic learning. This indicates that students' overall attitude toward the ease of use and usefulness of generative AI chatbots directly impacts their willingness to engage in collaborative learning activities using these platforms. Therefore, the study underscores the importance of addressing students', lecturers and other stakeholders' attitudes toward generative AI chatbots as part of efforts to promote its use for academic purposes (learning). By enhancing the user's perceptions of the ease of use and usefulness of generative AI chatbots, higher education institutions can positively influence students' attitudes and, in turn, their intention to utilize generative AI chatbots for learning purposes.

**Actual usage**

Actual usage refers to the real-world application and utilization of a technology or system by individuals in their daily activities. In the context of theoretical models such as the Model of Personal Computer Utilization (MPCU), actual usage represents the tangible behavior of individuals using technology in their work or personal tasks. Actual usage is a critical component in understanding the effectiveness of this conceptual framework in predicting and explaining generative AI chatbots acceptance and adoption in higher education. It provides insights into how students, lecturers, and other higher education stakeholders interact with generative AI chatbots in practical settings, shedding light on the factors that influence their ongoing use of the technology beyond initial acceptance.

In this study, studying actual usage allows students, researchers, lecturers, and other higher education stakeholders to assess the long-term impact of various factors such as attitudes on individuals' sustained engagement with generative AI chatbots. By examining actual usage patterns, higher education stakeholders can evaluate the predictive power and practical relevance of this conceptual model in capturing the complexities of human-technology interaction. Also, as insights into how these students, and lecturers use generative AI chatbots in their daily academic routines, higher education institutions can tailor their strategies to enhance user experience, productivity, and overall satisfaction.

**Recommendation and future study**

The key recommendations from the study on the adoption of generative AI chatbots in HE include:

1. **Developing a comprehensive framework:** The study recommends the development of a comprehensive framework for the use of generative AI chatbots in HEIs. The Generative AI Chatbots Acceptance Model (GAICAM) is proposed as a framework that incorporates aspects from known theories such as TAM, TRI 2.0, UTAUT2, revised TAM, and TPB to explain the acceptance of generative AI chatbots in HE.
2. **Addressing crucial components:** The study emphasizes the need to address crucial components such as readiness, perception, basic infrastructure, personal factors, attitude, and actual usage to promote a more seamless adoption process. It also emphasizes the importance of innovativeness and optimism in fostering early acceptance and support of AI chatbots in educational settings.
3. **Exploring the impact of AI chatbots on student learning outcomes:** Future research, according to the study, ought to look at how AI chatbots affect students' learning results over the long run. This could include exploring the effectiveness of generative AI chatbots in improving student engagement, academic performance, and retention rates.
4. **Understanding the role of AI chatbots in facilitating research and academic support:** The study recommends exploring the role of generative AI chatbots in facilitating research and academic support. This could include investigating how generative AI chatbots can be used to support faculty research, streamline administrative tasks, and provide personalized academic support to students.
5. **Examining cultural and contextual factors:** The study suggests that future research should examine the cultural and contextual factors impacting the acceptance of AI chatbots in diverse educational settings. This could include exploring how cultural norms, values, and beliefs impact the adoption and use of generative AI chatbots in different regions and countries.

The proposed future studies based on the recommendations include:

1. **Long-term impact studies:** Research that delves into the long-term impact of generative AI chatbots on student learning outcomes, including academic performance, engagement, retention, and research.
2. **Role of AI chatbots in academic support:** Studies focusing on the specific ways generative AI chatbots can support faculty research, streamline administrative tasks, and provide personalized academic support to students.

3. **Cultural and contextual studies:** Research that examines how cultural and contextual factors influence the acceptance and use of generative AI chatbots in diverse educational settings.

These recommendations and proposed future studies provide valuable guidance for researchers and educational institutions looking to integrate generative AI chatbots effectively into higher education.

**Contributions of the study**

The corpus of information on the topic is significantly expanded by the study on the use of generative AI chatbots in HE. First off, the paper provides a thorough framework for AI chatbot deployment in HEIs. By synthesizing elements from well-known frameworks such as the TAM, UTAUT2, revised TAM, and TPB, along with variables like optimism, innovativeness, discomfort, insecurity, and many others, the study provides a strategic roadmap for incorporating AI chatbots into educational systems. This framework strikes a balance between technological preparedness and user attitudes, addressing crucial components such as readiness, perception, basic infrastructure, personal factors, attitude, and actual usage. As a result, the study's framework offers valuable guidance for educational institutions seeking to successfully implement generative AI chatbots.

Secondly, the study identifies key components that influence the adoption of AI chatbots in HEIs. By exploring the possibilities and constraints of generative AI chatbots in HE, the study sheds light on factors such as technological preparedness, user attitudes, innovativeness, and optimism. This identification of key factors provides insights into the challenges and opportunities associated with the adoption of generative AI chatbots in educational settings, thereby contributing to a deeper understanding of the dynamics involved in integrating generative AI chatbots into higher education. Furthermore, the study explores the potential benefits of generative AI chatbots in HE, such as improved student engagement, streamlined administrative tasks, and personalized academic support. By highlighting these potential benefits, the study provides a compelling case for the adoption of generative AI chatbots in educational settings. This investigation of possible advantages adds to the expanding corpus of research on the effects of generative AI chatbots in HE by providing insights into the revolutionary potential of these tools in improving the learning environment for lecturers and students alike.

Lastly, the study suggests future research directions, such as investigating the long-term impact of generative AI chatbots on student learning outcomes, exploring the role of generative AI chatbots in facilitating research and academic support, and understanding the cultural and contextual factors influencing the acceptance of generative AI chatbots in diverse educational settings. By identifying these future research directions, the study provides a roadmap for researchers and practitioners interested in further exploring the adoption and impact of generative AI chatbots in educational settings. This contribution to the identification of future research directions serves to guide and inspire further scholarly inquiry into the evolving role of generative AI chatbots in HE, thereby enriching the academic knowledge in this field.

**Limitations of the study**

While the study on the adoption of generative AI chatbots in HE provides valuable insights, some limitations should be considered. These limitations include:

1. **Limited timeframe:** The study focuses on peer-reviewed English-language publications published between 2020 and 2023. This limited timeframe may exclude relevant studies published before or after this period.
2. **Limited scope:** The study focuses on the adoption of generative AI chatbots in HE institutions and does not consider other educational settings such as K-12 schools or vocational training centers.
3. **Limited geographical coverage:** The study primarily focuses on studies conducted in English-speaking countries, and the generalizability of the findings to other regions and countries can be limited.
4. **Limited methodological diversity:** The study primarily relies on a comprehensive analysis of extant literature from databases such as IEEE, ACM, ScienceDirect, and Google Scholar. While this approach provides a broad overview of the literature, it may exclude studies that use different research methods such as case studies or ethnographic research.
5. **Limited sample size:** The lack of information in the study about the sample sizes of the studies that were analyzed might restrict how broadly the results can be applied.

The study provides valuable insights into the adoption of generative AI chatbots in higher education, the limitations should be considered when interpreting the findings. Future research should address these limitations to provide a more comprehensive understanding of the adoption and impact of generative AI chatbots in educational settings.

**Conclusions**

This paper introduces the Generative AI Chatbots Acceptance Model (GAICAM), a comprehensive framework that incorporates aspects from known theories such as TAM, TRI 2.0, UTAUT2, revised TAM, and TPB to explain the acceptance of AI chatbots in higher education. The study recognizes the growing interest in and promises AI chatbots to change education, especially in higher education, by providing individualized support, facilitating research, and improving the quality of the learning process.

The GAICAM framework posits six crucial components that impact the acceptability and integration of artificial intelligence (AI) chatbots in higher education environments. These components include readiness, perception, basic infrastructure, personal factors, attitude, and actual usage. The study highlights how crucial it is to deal with these elements in order to promote a more seamless adoption process. For instance, innovativeness and optimism emphasize having a positive view of life and being open to embracing new technology, all of which are essential for the early acceptance and support of generative AI chatbots in educational settings. Contrarily, unease and insecurity stand in for difficulties or impediments to adoption, emphasizing the necessity of resolving issues with the intricacy, dependability, and security of generative AI chatbots.

A strong basis for the framework is provided by the research methodology, which comprises an extensive evaluation of the body of current literature and empirical data. The study's conclusions show that students and educators are becoming more aware of and interested in AI chatbots, but they also highlight adoption hurdles that must be removed. The GAICAM framework provides higher education institutions with a methodical way to handle the challenges involved in implementing generative AI chatbots. This paradigm intends to allow a broader and successful integration of generative AI chatbots in higher education, thereby improving the educational process and outcomes, by taking into account aspects that impact user adoption and resolving potential impediments. The study emphasizes how important it is to keep researching and adapting in this quickly changing industry to make sure generative AI chatbots are used as efficiently and effectively as possible in educational settings.

**References**

1. Adom, D., Hussein, E. K., & Agyem, J. A. (2018). Theoretical and conceptual framework: Mandatory ingredients of a quality research. *International journal of scientific research*, 7(1), 438-441.

2. Akgun, S., & Greenhow, C. (2022). Artificial intelligence in education: Addressing ethical challenges in K-12 settings. *AI and Ethics*, *2*(3), 431-440.

3. Alkhwaldi, A. F. A., & Kamala, M. A. (2017). Why do users accept innovative technologies? A critical review of models and theories of technology acceptance in the information system literature.

4. Amadu, L., Muhammad, S. S., Mohammed, A. S., Owusu, G., & Lukman, S. (2018). Using technology acceptance model to measure the ese of social media for collaborative learning in Ghana. *JOTSE*, *8*(4), 321-336.

5. Barbaranelli, C., Farnese, M. L., Tramontano, C., Fida, R., Ghezzi, V., Paciello, M., & Long, P. (2018). Machiavellian ways to academic cheating: A mediational and interactional model. *Frontiers in Psychology*, *9*, 695.

6. Brown, S. A., & Venkatesh, V. (2005). Model of adoption of technology in households: A baseline model test and extension incorporating household life cycle. *MIS Quarterly*, 29(3), 399-426.

7. Chan, C. K. Y. (2023). A comprehensive AI policy education framework for university teaching and learning. *International journal of educational technology in higher education*, *20*(1), 38.

8. Chukwuere, J. E. (2021). Theoretical And Conceptual Framework: A Critical Part of Information Systems Research Process and Writing. *Review of International Geographical Education Online*, *11*(9).

9. Chukwuere, J. E., Ntseme, O. J., & Shaikh, A. A. (2021). Toward the development of a revised technology acceptance model. In *Proceedings of the International Conference on Electronic Business*. International Consortium for Electronic Business.

10. Davis, F. D. (1989). Perceived usefulness, perceived ease of use, and user acceptance of information technology. *MIS quarterly*, 319-340.

11. Dejene, W. (2021). Academic cheating in Ethiopian secondary schools: Prevalence, perceived severity, and justifications. *Cogent Education*, *8*(1), 1866803.

12. Gimpel, H., Hall, K., Decker, S., Eymann, T., Lämmermann, L., Mädche, A., ... & Vandrik, S. (2023). *Unlocking the power of generative AI models and systems such as GPT-4 and ChatGPT for higher education: A guide for students and lecturers* (No. 02-2023). Hohenheim Discussion Papers in Business, Economics and Social .Sciences.

13. Gupta, A., Hathwar, D., & Vijayakumar, A. (2020). Introduction to AI chatbots. *International Journal of Engineering Research and Technology*, *9*(7), 255-258.

14. Hasal, M., Nowaková, J., Ahmed Saghair, K., Abdulla, H., Snášel, V., & Ogiela, L. (2021). Chatbots: Security, privacy, data protection, and social aspects. *Concurrency and Computation: Practice and Experience*, *33*(19), e6426.
15. Ilieva, G., Yankova, T., Klisarova-Belcheva, S., Dimitrov, A., Bratkov, M., & Angelov, D. (2023). Effects of Generative Chatbots in Higher Education. *Information*, *14*(9), 492.
16. Kooli, C. (2023). Chatbots in education and research: A critical examination of ethical implications and solutions. *Sustainability*, *15*(7), 5614.
17. Labadze, L., Grigolia, M., & Machaidze, L. (2023). Role of AI chatbots in education: systematic literature review. *International Journal of Educational Technology in Higher Education*, *20*(1), 56.
18. Liao, S., Hong, J. C., Wen, M. H., & Pan, Y. C. (2018). Applying technology acceptance model (TAM) to explore users' behavioral intention to adopt a performance assessment system for E-book production. *EURASIA Journal of Mathematics, Science and Technology Education*, *14*(10), em1601.
19. Liu, L., Subbareddy, R., & Raghavendra, C. G. (2022). AI Intelligence Chatbot to Improve Students Learning in the Higher Education Platform. *Journal of Interconnection Networks*, *22*(Supp02), 2143032.
20. Marikyan, M., & Papagiannidis, P. (2021). Unified theory of acceptance and use of technology. *TheoryHub book*.
21. Maruping, L. M., Bala, H., Venkatesh, V., & Brown, S. A. (2017). Going beyond intention: Integrating behavioral expectation into the unified theory of acceptance and use of technology. *Journal of the Association for Information Science and Technology*, *68*(3), 623-637.
22. Miao, J., Thongprayoon, C., Suppadungsuk, S., Garcia Valencia, O. A., Qureshi, F., & Cheungpasitporn, W. (2023). Ethical Dilemmas in Using AI for Academic Writing and an Example Framework for Peer Review in Nephrology Academia: A Narrative Review. *Clinics and Practice*, *14*(1), 89-105.
23. Momonov, G., & Mirtskhulava, L. (2021). Artificially intelligent chatbots for higher education: A review of empirical literature. https://www.researchgate.net/publication/361856793_Artificially_intelligent_chatbots_for_higher_education_A_review_of_empirical_literature
24. Ntseme, O. J. (2019). *Investigating e-health readiness of higher education institution students* (Doctoral dissertation, North-West University (South Africa)).

25. Okonkwo, C. W., & Ade-Ibijola, A. (2021). Chatbots applications in education: A systematic review. *Computers and Education: Artificial Intelligence*, p. 100033.
26. Oniani, D., Hilsman, J., Peng, Y., Poropatich, R. K., Pamplin, J. C., Legault, G. L., & Wang, Y. (2023). Adopting and expanding ethical principles for generative artificial intelligence from military to healthcare. *NPJ Digital Medicine*, *6*(1), 225.
27. Ou, A. W., Stöhr, C., & Malmström, H. Academic Communication with Ai-Powered Language Tools in Higher Education: From a Post-Humanist Perspective. *Available at SSRN 4589865*.
28. Parasuraman, A. (2000). Technology Readiness Index (TRI) a multiple-item scale to measure readiness to embrace new technologies. *Journal of service research*, *2*(4), 307-320.
29. Park, E. J., Park, S., & Jang, I. S. (2013). Academic cheating among nursing students. *Nurse education today*, *33*(4), 346-352.
30. Rasul, T., Nair, S., Kalendra, D., Robin, M., de Oliveira Santini, F., Ladeira, W. J., ... & Heathcote, L. (2023). The role of ChatGPT in higher education: Benefits, challenges, and future research directions. *Journal of Applied Learning and Teaching*, *6*(1).
31. Sandu, N., & Gide, E. (2019, September). Adoption of AI-Chatbots to enhance student learning experience in higher education in India. In *2019 18th International Conference on Information Technology Based Higher Education and Training (ITHET)* (pp. 1-5). IEEE.
32. Son, M., & Han, K. (2011). Beyond the technology adoption: Technology readiness effects on post-adoption behavior. *Journal of Business Research*, *64*(11), 1178-1182.
33. Su, J., & Yang, W. (2023). Unlocking the power of ChatGPT: A framework for applying generative AI in education. *ECNU Review of Education*, 20965311231168423.
34. Tlou, E. R. (2009). *The application of the theories of reasoned action and planned behaviour to a workplace HIV/AIDS health promotion programme* (Doctoral dissertation, University of South Africa).
35. Tsikriktsis, N. (2004). A technology readiness-based taxonomy of customers: A replication and extension. *Journal of service research*, *7*(1), 42-52.
36. Venkatesh, V., Thong, J. Y., & Xu, X. (2016). Unified theory of acceptance and use of technology: A synthesis and the road ahead. *Journal of the Association for Information Systems*, 13(1), 1-70.
37. Williams, R. T. (2024, January). The ethical implications of using generative chatbots in higher education. In *Frontiers in Education* (Vol. 8). Frontiers Media SA.

38. Yang, S., & Evans, C. (2019, November). Opportunities and challenges in using AI chatbots in higher education. In *Proceedings of the 2019 3rd International Conference on Education and E-Learning* (pp. 79-83).
39. Yu, H. (2024). The application and challenges of ChatGPT in educational transformation: New demands for teachers' roles. *Heliyon*.